\documentclass[prl,aps,twocolumn]{revtex4} 
\usepackage{psfig}

\begin{document}

\title{On the definition of temperature in dense granular media} 
 
\author{Vittoria Colizza$^1$, Alain Barrat$^2$ and Vittorio Loreto$^1$}

\affiliation{$^1$ Universit\`a degli Studi di Roma ``La Sapienza'',
Dipartimento di Fisica, P.le A. Moro 5, 00185 Rome, Italy and INFM,
Unit\`a di Roma 1 \\ $^2$ Laboratoire de Physique Th{\'e}orique Unit{\'e}
Mixte de Recherche UMR 8627, B{\^a}timent 210, Universit{\'e} de
Paris-Sud, 91405 Orsay Cedex, France }
 
\date{\today}

\begin{abstract}
In this Letter we report the measurement of a pseudo-temperature for
compacting granular media on the basis of the Fluctuation-Dissipation
relations in the aging dynamics of a model system. From the violation
of the Fluctuation-Dissipation Theorem an effective temperature
emerges (a dynamical temperature $T_{dyn}$) whose ratio with the
equilibrium temperature $T_d^{eq}$ depends on the particle density. We
compare the results for the Fluctuation-Dissipation Ratio (FDR)
$T_{dyn}/T_d^{eq}$ at several densities with the outcomes of Edwards'
approach at the corresponding densities. It turns out that the FDR and
the so-called Edwards' ratio coincide at several densities (very
different ages of the system), opening in this way the door to
experimental checks as well as theoretical constructions.(PACS:
05.70.Ln, 05.20.-y, 45.70.Cc)
\end{abstract} 

\maketitle

The study of compact granular matter through statistical physics tools
is the subject of a sustained interest~\cite{exp}. Granular media enter only
partially in the framework of equilibrium statistical mechanics and
their dynamics constitutes a very complex problem of non-equilibrium,
which poses novel questions and challenges to theorists and
experimentalists. The very possibility to construct a coherent
statistical mechanics for these systems is still matter of debate,
although everybody agrees that such an approach, if possible,
would allow for a much deeper and global understanding of the problem.

One of the main obstacles in this direction is the non-thermal
character of these systems: thermal energy is so negligibly small with
respect to other energy contributions (e.g. potential energy) that for
all the practical purposes these systems live virtually at zero
temperature. One of the most important consequences is that, unless
perturbed in some way (e.g. driving energy into the system), a
granular system cannot explore spontaneously its phase space but it
remains trapped in one of the numerous metastable configurations.
Understanding the structure of the phase space which is left invariant
by the dynamics is then crucial for the construction of a
thermodynamical description of these non-thermal systems.

A very ambitious approach, in this direction, has been put forward by
S. Edwards and co-workers~\cite{edwards,anita}, by proposing an
equivalent of the microcanonical ensemble: macroscopic quantities in a
jammed situation should be obtained by a flat average over all {\em
blocked configurations} (i.e. in which every grain is unable to move)
of given volume, energy, etc... The strong assumption here is that all
blocked configurations are treated as equivalent and have the same
weight in the measure. This approach, based on the idea of describing
granular material with {\em a small number of parameters}, leads to
the introduction of an entropy $S_{edw}$, given by the logarithm of
the number of blocked configurations of given volume, energy, etc.,
and its corresponding density $s_{edw}\equiv S_{edw}/N$.  Associated
with this entropy are the state variables such as `compactivity'
${\cal{X}}^{-1}=\frac{\partial}{\partial V}S_{edw}(V)$ and `temperature'
${{T}}^{-1}=\frac{\partial}{\partial E}S_{edw}(E)$.

Very recently, important progresses in this direction have been
reported in various contexts: a tool to systematically construct
Edwards' measure, defined as the set of blocked configurations of a
given model, was proposed in~\cite{bakulose1,bakulose2}; it was used
to show that the outcome of the aging dynamics of the Kob-Andersen
model (a kinetically constrained lattice gas model) was correctly
predicted by Edwards' measure. Moreover, the validity and relevance of
Edwards' measure have been demonstrated for one-dimensional
phenomenological models~\cite{brey}, for spin models with ``tapping''
dynamics~\cite{Dean}, and for sheared hard spheres~\cite{Makse}.

In this Letter we focus on the definition of a pseudo-temperature for
granular media on the basis of the Fluctuation-Dissipation relations
in the out-of-equilibrium, aging, dynamics of a model
system~\cite{prl}, and on its relation to Edwards' measure.  From the
violation of the Fluctuation-Dissipation Theorem an effective
temperature emerges (from now onward indicated as dynamical
temperature $T_{dyn}$) whose ratio with the equilibrium temperature
$T_d^{eq}$ depends on the particle density. We compare the results for
the Fluctuation-Dissipation Ratio (FDR) $T_{dyn}/T_d^{eq}$ at several
densities with the outcomes of Edwards' approach at the corresponding
densities. It turns out that the FDR and the so-called Edwards' ratio
coincide at several densities (very different ages of the system),
opening in this way the door to experimental checks as well as
theoretical constructions.

It is interesting to mention recent approaches that are complementary
to ours.  On the one hand, in the context of aging supercooled
liquids, the inherent structure strategy does not address the question
of a determination of a static distribution, but promising results
show that this measure, if it exists, is insensitive to the details of
the thermal history ~\cite{kst}. The link between this strategy and
Edwards' measure has been discussed in~\cite{bakulose2,fierro}. On
the other hand, recent works on the possibility of a dynamical
definition of temperature have focused on sheared, stationary 
systems~\cite{ludo2,liu}. These studies are clearly
complementary to the present one, which addresses the problem
of the relation between the
value of $T_{dyn}$ and a static measure, in aging (non-stationary) systems.

The model we consider is a version of the ``Tetris'' model~\cite{prl},
which has been shown to reproduce several features of granular media
like aging~\cite{nico,bl}, memory~\cite{bl2},
self-structuring~\cite{self-stru} etc.  In the framework of this model,
some of us have already provided one of the first evidences of the
validity of Edwards' measure~\cite{bakulose2}.

We focus in particular on an homogeneous system with no preferential
direction in order to avoid any kind of instability or large-scale
structure formation~\cite{self-stru}. The case with gravity which
imposes a preferential direction will be discussed
elsewhere~\cite{inprep}.  In the version of the model we use,
``T''-shaped particles diffuse on a square lattice, with the only
constraint that no superposition is allowed: for two nearest-neighbor
particles, the sum of the arms oriented along the bond connecting the
two particles has to be smaller than the bond length (for each
particle, the three arms of the `T'' have length $\frac{3}{4} d$,
where $d$ sets the bond size on the square lattice). The maximum
density allowed is then $\rho_{max}=2/3$.  This model represents a
clear example of a non-thermal system. The Hamiltonian is zero and the
temperature itself is therefore irrelevant at equilibrium, only its
ratio with an imposed chemical potential being important.

The out-of-equilibrium compaction dynamics without gravity is
implemented as follows: starting from an empty lattice, particles are
randomly deposited, without diffusion and without violation of the
geometrical constraints. This random sequential absorption process
yields a reproducible initial density of $\rho \approx
0.547$. Alternating diffusions and additions of particles are then
attempted, allowing to increase the density of the system, which
remains homogeneous in the process. In order to overcome the problem
related to the simulation of slow processes and obtain a reasonable
number of different realizations to produce clean data, we have
devised a fast algorithm (in the spirit of Bortz-Kalos-Lebowitz
algorithm~\cite{bkl}) where the essential ingredient is the updating
of a list of mobile particles (whose number is $n_{mob}$). At each
time step one selects a mobile particle and move it in the direction
chosen only if no violation of the constraints occurs. The time is
incremented of an amount $\Delta t = 1/ n_{mob}$.  We refer to
~\cite{inprep} for the details of the fast algorithm.

We have simulated lattices of linear size $L=50,\;100,\;200$, in order to
ensure that finite-size effects were irrelevant. We have chosen
periodic boundary conditions on the lattice, having verified that
other types of boundary conditions (e.g closed ones) gave the same
results.

During the compaction, we monitor the following quantities: the
density of particles $\rho(t)$, the density of mobile particles
$\rho_{mob}(t)$. Moreover in order to establish the Fluctuation-Dissipation
relations we measure the mean square displacement $B(t+t_w,t_w)$ and the
integrated response function  $\chi(t+t_w,t_w)$.
The mean square displacement is defined as  
\begin{displaymath}
B(t+t_w,t_w)\,=\,\frac{1}{N}\,\sum_{i=0}^{N}\,\sum_{r=x,y}\,\langle\,
\left[ r_i(t+t_w,t_w)-r_i(t_w) \right]^2 \rangle 
\nonumber
\end{displaymath}
where $N$ is the number particles present in the system
at time $t_w$,  $r_i$ is the coordinate ($x$ or $y$) 
of the $i$-th particle and the brackets $\langle \rangle$ 
indicate the average over several realizations.

In order to measure the integrated response function
$\chi(t+t_w,t_w)$, we make a copy of the system at time $t_w$ and
apply to it a small random perturbation, varying the diffusion
probability of each particle from $p=\frac{1}{4}$ to
$p^{\epsilon}=\frac{1}{4} + f_i^r \cdot \epsilon$, where $f_i^r=\pm1$
is a random variable associated to each grain independently for each
possible direction ($r=x,y$), and $\epsilon$ represents the
perturbation strength. For a constant field we obtain (see also
~\cite{bakulose1,bakulose2,inprep}):
\begin{displaymath}
\chi(t+t_w,t_w)\,=\,\frac{1}{2 \epsilon N}
\sum_{i=1}^N \sum_{r=x,y} \langle f_i^r
\cdot \Delta r_i (t+t_w) \rangle
\nonumber
\end{displaymath}
where $\Delta r_i (t+t_w)=r_i^{repl}(t+t_w)-r_i(t+t_w)$ is the
difference between the displacements taking place in the two
systems evolving with the same succession of random numbers. In this
letter we present the results obtained with a perturbation strength
$\epsilon=0.005$, having checked that for $0.002< \epsilon <0.01$
non-linear effects are absent.

If the system was in equilibrium we would expect $B$ and $\chi$ to be
linearly related by
\begin{equation}
\chi(t+t_w,t_w) = \frac{X_{dyn}}{T^{eq}_d} B(t+t_w,t_w),
\label{FDT}
\end{equation}
where $X_{dyn}$ is the so-called Fluctuation-Dissipation Ratio (FDR)
which is unitary in equilibrium. Any deviations from this linear law
signals a violation of the Fluctuation-Dissipation Theorem
(FDT). Nevertheless it has been shown, first in mean-field
models~\cite{CuKu}, then in various
simulations~\cite{parisi,barratkob} how in several aging systems
violations from (\ref{FDT}) reduce to the occurrence of two regimes: a
quasi-equilibrium regime with $X_{dyn}=1$ (and time-translation
invariance) for ``short'' time separations ($t \ll t_w$), and the
aging regime with $X_{dyn} \le 1$ for large time separations. This
second slope is typically referred to as a dynamical temperature
$T_{dyn} \ge T^{eq}_d$ such that $X_{dyn} =
T^{eq}_d/T_{dyn}$~\cite{CuKuPe}.

In our case, as the density increases, an aging behaviour is obtained:
the system falls out of equilibrium and $\rho_{mob}(t)$ gets smaller
than the corresponding value at equilibrium~\cite{bakulose2}. Accordingly,
violations of (\ref{FDT}) are expected.

If the compaction process is stopped at a certain time $t_w$, the
system relaxes toward equilibrium and one obtains a time-translation
invariant behaviour for $\chi$ and $B$; this is the so-called regime
of interrupted aging which features an increase of $\rho_{mob}$ to its
equilibrium value and a single linear relation for the $\chi$ vs. $B$
parametric plot. These measures therefore give us the value of the
equilibrium fluctuation-dissipation ratio, $T_d^{eq}$, which does
actually not depend on the density reached at $t_w$.

\begin{figure}[htb]
\centerline{
        \psfig{figure=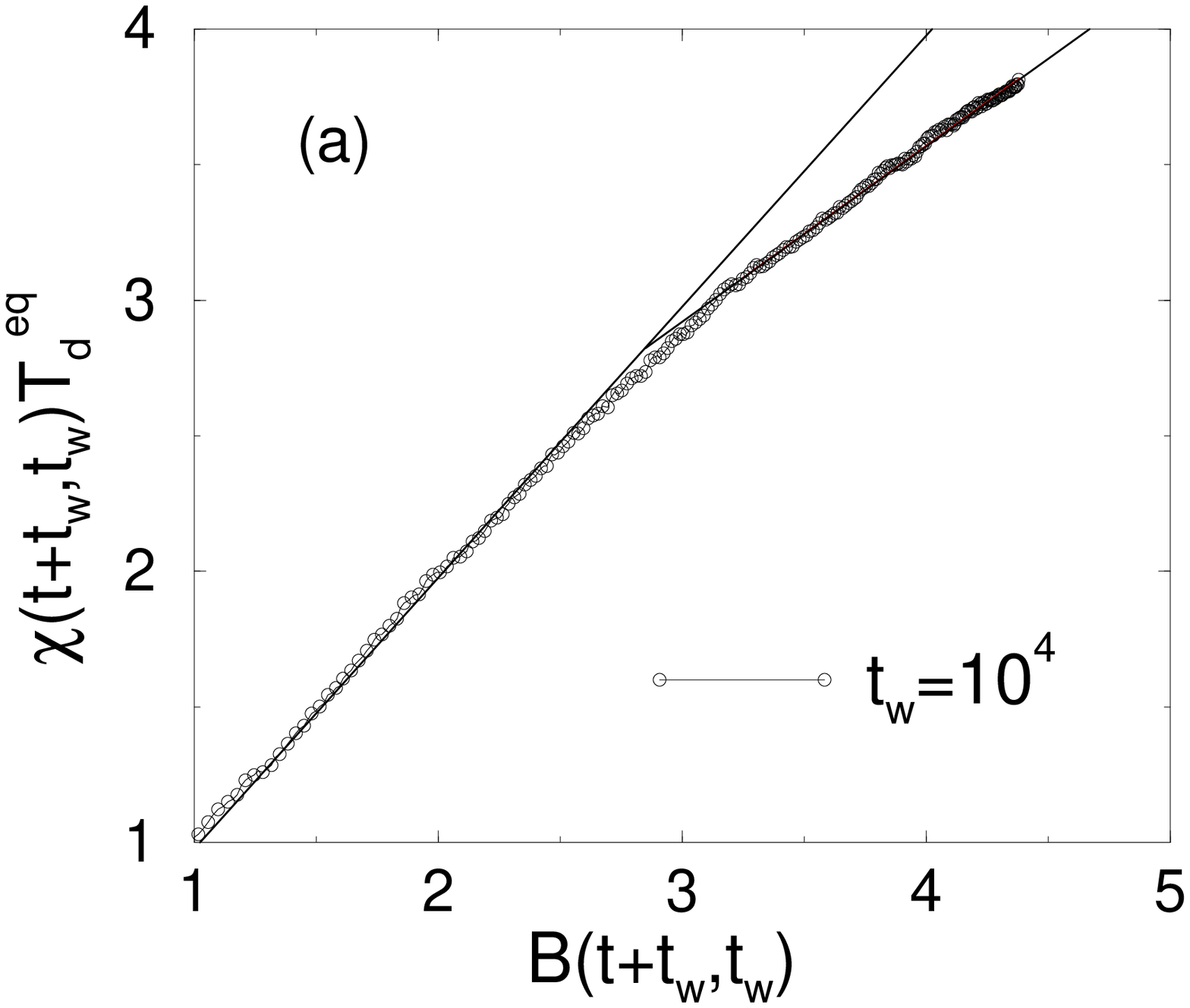,width=4.3cm,angle=0}
	\psfig{figure=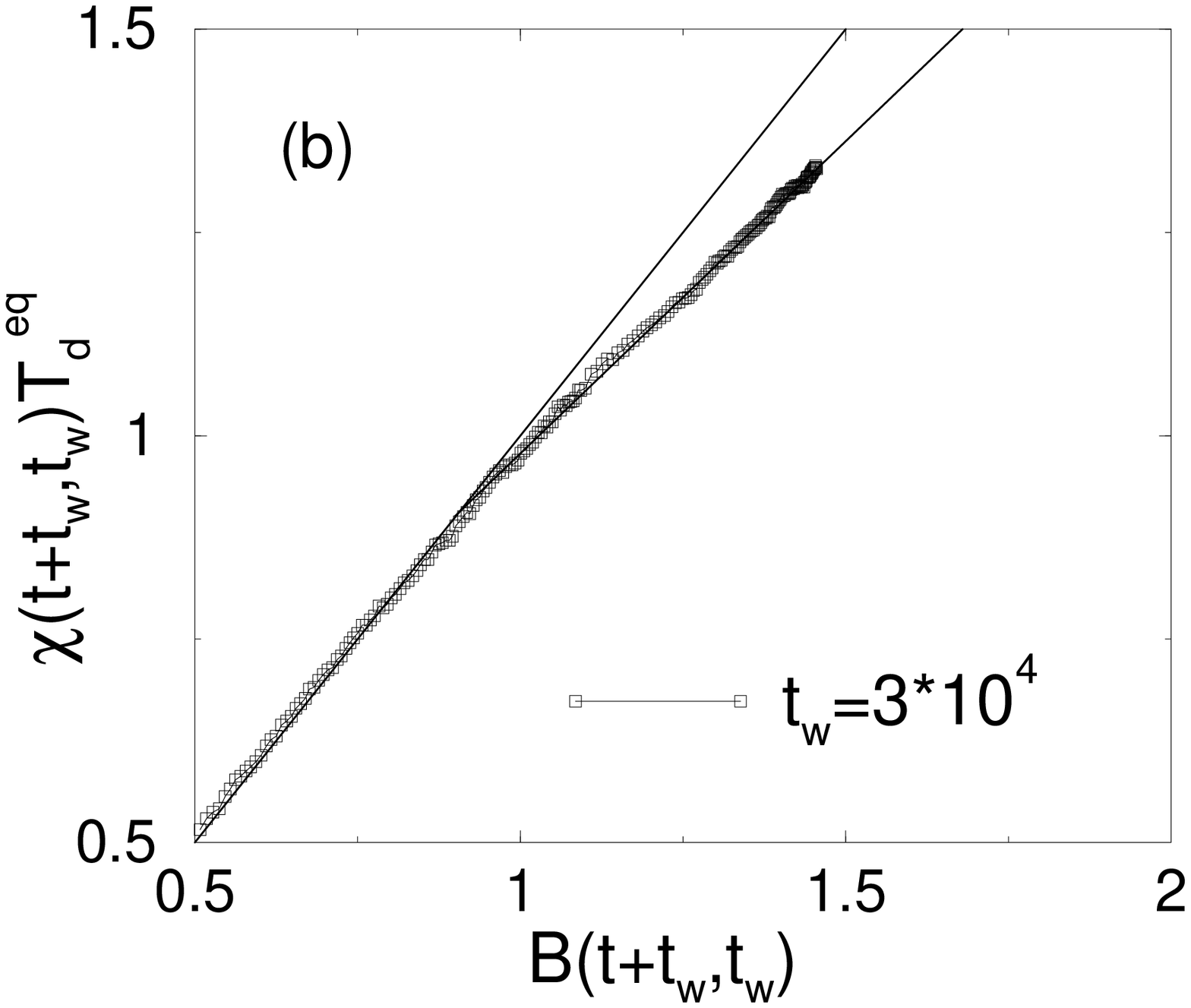,width=4.3cm,angle=0}}
\caption{Einstein relation in the Tetris model: plot of the mobility
$\chi(t_w+t,t_w) T_d^{eq}$ vs. the mean-square displacement $B(t_w+t,t_w)$, for
$t_{w}^1=10^4$ and $t_{w}^2=3\cdot 10^4$. The unitary slope of the full straight
line corresponds to the equilibrium case, obtained
for the dynamics at constant density (interrupted aging).}
\label{fig1}
\end{figure}

If, on the other hand, the compaction process is not stopped, the
system features an aging behaviour and the $\chi$ vs. $B$ parametric
plots, displayed in figure \ref{fig1}, show two different linear
behaviours: after a first quasi-equilibrium regime in which the plot
has slope $T_d^{eq}$, a violation of FDT is observed, with the
existence of a dynamical temperature $T_{dyn}$ which depends on $t_w$
and therefore on the density. We obtain in particular $X_{dyn}^1 =
0.646 \pm 0.002$, $X_{dyn}^2 = 0.767 \pm 0.005$, $X_{dyn}^3 = 0.784
\pm 0.005$ at $t_{w}^1=10^4$, $t_{w}^2=3 \cdot 10^4$, and $t_{w}^3=5
\cdot 10^4$.

\begin{figure}[htb]
\centerline{
        \psfig{figure=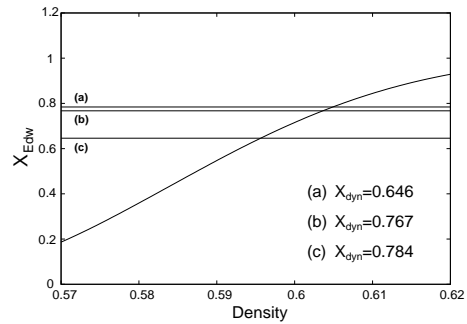,width=6cm}}
\caption{Edwards' ratio $X_{Edw}$ as a function of density. The
horizontal lines correspond to the dynamical ratios $X_{dyn}$,
measured (from bottom to top) at $t_w=10^4,\ 3\cdot10^4,\ 5 \cdot
10^4$ and determine the values $\rho_1 \approx 0.596$, $\rho_2 \approx
0.603$, $\rho_3 \approx 0.605$, to be used in the comparison with the
results reported in Fig. 3.}
\label{fig:comp} 
\end{figure}

We are now able to compare the values of the dynamical
measures with the outcome of Edwards' measure.  The construction of
the equilibrium and Edwards' measures has been described in
~\cite{bakulose2}. In particular, an efficient sampling of the blocked
configurations, and therefore Edwards' measure, is obtained by the use
of an auxiliary model whose energy is defined to be the number of
mobile particles: the introduction of an auxiliary temperature and an
annealing procedure then yields configurations with no mobile
particles.  Having computed equilibrium and Edwards' entropies as a
function of density, as in~\cite{bakulose1,bakulose2}, we measure the
ratio of the slopes (that we denote as Edwards' ratio)
\begin{equation}
X_{Edw} = \frac{ds_{Edw}(\rho)}{d\rho} \Bigg/
	\frac{ds_{equil}(\rho)}{d\rho},
\end{equation}
which is plotted in figure (\ref{fig:comp}).  This ratio approaches
$1$ as $\rho \to 2/3$, since at the maximum density all configurations
become blocked and therefore equilibrium and Edwards' measures become
equivalent.

\begin{figure}[htb]
\centerline{
        \psfig{figure=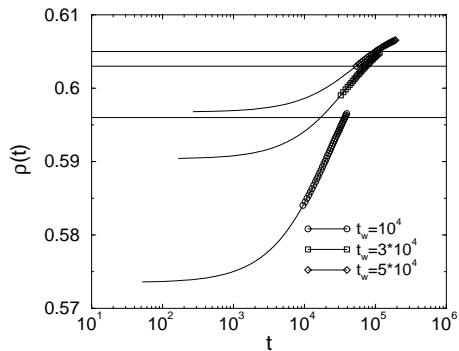,width=6cm,angle=0}}
\caption{Evolution of the density during the measurements of $\chi$
and $B$, for $t_w=10^4,\; 3 \cdot 10^4,\; 5 \cdot 10^4$. The evolution during the
quasi-equilibrium part is plotted with lines, and during the violation
of FDT with symbols. The horizontal lines correspond to the densities
$\rho_1$, $\rho_2$, $\rho_3$ from Fig 2.}
\label{fig:rho}
\end{figure}

The values of the dynamical Fluctuation-Dissipation ratio are also
reported in the same figure (with no error bars since they are too
small) and yield the following densities: $\rho_1 \approx 0.596$ for
$t_{w}^1=10^4$, $\rho_2 \approx 0.603$ for $t_{w}^2=3. 10^4$, $\rho_3
\approx 0.605$ for $t_{w}^3=10^5$.  On the other hand, the evolution
of the density of the system during the measurements is reported in
figure (\ref{fig:rho}).

\begin{table}[hbt]
\begin{center}
\begin{tabular}{lccc}
\hline
$$ & \mbox{\bf{ $X_{dyn}$ }} & \bf{ density } & \bf{density
interval}\\ \hline \bf{$t_w=10^4$} & $0.646$ & $0.596$ &
$[0.584,0.597]$\\ \hline \bf{$t_w=3\cdot10^4$} & $0.767$ & $0.603$ &
$[0.599,0.605]$\\ \hline \bf{$t_w=5\cdot10^4$} & $0.784$ & $0.605$ &
$[0.603,0.606]$\\ \hline
\end{tabular}
\caption{Fluctuation-Dissipation Ratio ($X_{dyn}$) as obtained from
numerical data fits with the corresponding densities obtained from
figure (\ref{fig:comp}). The last column reports the density intervals
explored during the compaction dynamics (starting from the times where
deviations from equilibrium become evident) for
$t_w=10^4,\,3\cdot10^4,\,5\cdot10^4$.}
\end{center}
\end{table}

Since the measurements are performed {\em during} the compaction, the
density is evolving, going from $\rho(t_w)$ to $\rho(t_w+t_{max})$. In
each case, we obtain that indeed $\rho_i \in [\rho(t_{w}^i),
\rho(t_{w}^i +t_{max})]$ where we have denoted with $\rho_i$ the
densities obtained from figure (\ref{fig:comp}) for different values
of $t_w$ ($i=1,2,3$).  More precisely, $\rho_i$ is very close to
$\rho(t_{w}^i +t_{max})$. This is to be expected since the measure of
the FDT violation is made for times much larger than $t_{w}^i$ and,
since the compaction is logarithmic, the system spends actually more
time at densities close to $\rho(t_{w}^i +t_{max})$ than to
$\rho(t_{w}^i)$.

The validation of Edwards' hypothesis in various model systems has
made important steps forwards recently; here we have focused, for a
model with only geometrical constraints, on the definition of a
dynamical temperature and on its link with Edwards' measure. While the
density increases, the measured dynamical temperature decreases,
following closely the ratio between the equilibrium and the Edwards'
entropies, the latter being obtained through a flat sampling of
blocked configurations.  While Edwards' proposal is supposed to be
only valid asymptotically, i.e. when one-time quantities are almost
stationary, our study clearly shows that it actually yields good
results even in a pre-asymptotic regime, when the density is still
evolving a lot.

Two remarks are in order. It is interesting to investigate the limits
of validity of Edwards' approach. Though in this letter we have shown
that Edwards' measure works nicely in the idealized case of an
homogeneously compacting system, it is important to consider the case
of a compacting system under gravity, i.e. with a preferential
direction.  We shall report about this in~\cite{inprep}.  Another
crucial point concerns the fact that the dynamically defined effective
temperature could a priori depend on the observables used for the
measure of Fluctuation-Dissipation relations.  Its interpretation as a
temperature in a thermodynamical sense would then be questioned. While
the effective temperature is known to be observable independent for
mean-field models, this question has been addressed only recently in
realistic models: two recent studies on sheared (stationary)
thermal~\cite{ludo2} or athermal~\cite{liu} systems show the
consistency of various definitions~\cite{remark_shear}.  Our results
complement these studies by allowing to relate the value of $T_{dyn}$
to a static measure. Moreover, the observable-independence character
of $T_{dyn}$ in our case will be checked in~\cite{inprep}, by
considering systems consisting of 2 types of particles. In this
respect, the presence of a preferential direction could imply
limitations; for sheared, stationary systems, a preferential direction
exists, but only measures along orthogonal directions have been
performed~\cite{Makse,Ludo,ludo2,liu}.  {\large Acknowledgments} This
work has been partially supported by the European Network-Fractals
under contract No. FMRXCT980183.


\begin{thebibliography}{99}
\bibitem{remark_shear}
The models considered in~\cite{ludo2,liu} are sheared, and therefore
present a preferred direction. However the observables under investigation
concern only the other directions.
\end{thebibliography}
\end{document}